\documentclass[aps,prl,twocolumn,epsf,psfig,superscriptaddress]{revtex4}
\usepackage[dvips]{graphicx}
\draft
\begin{document}
\newcommand{\bea}{\begin{equation}}
\newcommand{\ber}{\begin{eqnarray}}
\newcommand{\eea}{\end{equation}}
\newcommand{\eer}{\end{eqnarray}}
\newcommand{\ct}{\cite}
\newcommand{\bi}{\bibitem}
\title
{Universality in the merging dynamics of parametric active contours: a study in
MRI-based lung segmentation}
\author{Amit K. Chattopadhyay}
\affiliation{
Max Planck Institute for the Physics of Complex Systems,
Noethnitzer Strasse 38,
01187 Dresden, Germany
}

\email{akc@mpipks-dresden.mpg.de}

\author{Nilanjan Ray}
\affiliation{
Department of Electrical and Computer Engineering, 
Thornton Hall, 
351 McCormick Road, 
P.O. Box 400743,  
Charlottesville, VA 22904-4743, U.S.A.} 
\email{nray@virginia.edu, acton@virginia.edu}
\author{Scott T. Acton}

\affiliation{
Department of Electrical and Computer Engineering, 
Thornton Hall, 
351 McCormick Road, 
P.O. Box 400743,  
Charlottesville, VA 22904-4743, U.S.A.}
\email{acton@virginia.edu} 
\date{\today}
%

%

\begin{abstract}
\noindent
Measurement of lung ventilation is one of the most reliable 
techniques of diagnosing pulmonary
diseases. The time consuming and bias prone traditional methods using 
hyperpolarized H${}^{3}$He and ${}^{1}$H magnetic resonance imageries have 
recently been improved by an automated technique based on multiple active 
contour evolution. Mapping results from an equivalent thermodynamic model, 
here we analyse the fundamental dynamics orchestrating the active contour (AC)
method. We show that the numerical method is inherently connected 
to the universal scaling behavior of a classical nucleation-like dynamics. 
The favorable comparison of the exponent values with the theoretical 
model render further credentials to our claim. \\

\noindent
PACS: 87.10. +e, 87.68. +z, 82.60. Nh
\end{abstract}

\pacs{87.10. +e, 87.68. +z, 82.60. Nh}

\maketitle

\noindent

Ventilation analysis is an authentic way of diagnosing lung airway diseases.
The ratio of the volume of ventilated (functional) portions of lungs
to the total lung volume is known as lung ventilation, which is
used in validating pulmonary drugs \cite{Ray}. The process involves 
two complementary 
magnetic resonance (MR) imaging modalities, the hyperpolarized 
helium-3 ($\rm{H}^{3}$He) imagery and the proton (${}^{1}$H) imagery. Lung 
functionality, including the volume of ventilated lungs, can be obtained from 
the former modality while lung anatomic details, including total lung volume, 
are accessed through the latter \cite{Ray}. Since manual investigation of
the MR 
imagery to compute lung ventilation is extremely time consuming,  
an {\em active contour (AC)} or {\em snake} based 
automated method has been proposed \cite{Ray,Xu} to compute the total 
lung volume from proton MR
imagery on a 
2-D slice-by-slice basis \cite{Ray}. A snake is defined as a massless 2-D thin
string 
(closed or open) that can move on the image domain driven by  
two types of forces, – internal elastic forces and external image
forces 
\cite{Kass}. Under the influence of these two forces an initial
contour (snake) clings to 
image edges and delineates an object. A snake always gives continuous
edges unlike any traditional edge detector (such as the Canny method 
\cite{Canny}), thereby eliminating any post-processing steps to
connect the 
detected broken edges. These two properties are particularly useful
when the 
object outline is broken and noisy as in most of the ${}^{1}$H MR
imagery. 
The snake method of finding the object boundary relies on the initial 
snake placement inside the image. If a small initial snake is placed 
inside a lung cavity on the MR image, while growing, the snake may be 
stopped by the associated numerical artifacts and may not capture the 
actual lung outline 
\cite{Ray}. Starting with a larger snake may result in missing the lung
cavity completely. A possible solution is to start
with multiple non-overlapping small snakes inside the lung cavity
and evolve (grow) them until they merge with each other and capture 
the cavity outline \cite{Ray}. During such a process, the growing 
snakes merge with each other into a single contour. This automatic
merging of non-overlapping snakes is characterized by certain attributes:
a) during the evolution process no 
two snakes overlap with each other, b) every snake stops evolving at
the object edge as a single snake does during its course of evolution,
and c) growing convex shaped snakes (e.g., circular or rectangular 
snakes) inside a convex object recovers the object boundary. 
Although merging of multiple 
snakes is experimentally verified in a multitude of cases,  
a concrete theoretical understanding of 
this merging snake approach remains a challenge. 

The aim of this letter 
is to bridge this gap by showing that the underlying principle of 
multiple snake-merging is governed by a {\em universal power law} behavior 
which originates from an inherent {\em nucleating} structure.  
From a biomedical engineering perspective, the study of the efficacy of the
lung cavity delineation method is crucial for robust clinical application. 
The lung cavity segmentation by merging snake method 
involves three steps: a) initially small 
non-overlapping contours are placed inside the lung cavity, b)
generalized gradient vector flow (GGVF) fields \cite{Xu} are
computed with a 
Dirichlet boundary condition on the initial circular snakes 
\cite{Ray}, and c) all the snakes are evolved simultaneously and 
independently of each other with the GGVF force field as the 
external force for the snakes. This automated lung cavity 
segmentation is attractive for a number of reasons. While 
other merging snake algorithms, such as the one proposed by 
McInerney and Terzopoulos \cite{McInerney} is computationally 
non-trivial compared to the original snake evolution algorithm 
of Kass et al. \cite{Kass}, the merging snake algorithm by 
Ray {\em et al}. maintains the same computational simplicity of 
Kass {\em et al}. algorithm. Also, based on the position 
of an initial snake, the rigidity parameters of the snakes can be
varied, so that on one hand delicate high curvature features, such as 
costophrenic angles, can be accurately captured, 
while on the other hand, snakes can be made sufficiently stiff in order 
to avoid capturing artifacts. Fig. 1 shows multiple snake 
initialization, evolution, merging and delineation of lung boundary 
by the Ray {\em et al'l} method.

\begin{figure}
\includegraphics[height=4.0cm,width=8.0cm,angle=0]{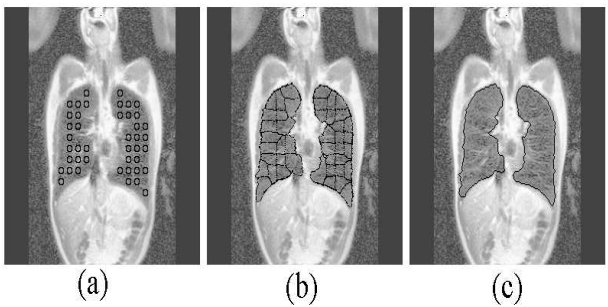}
\caption{Evolution of multiple snakes in a 2-D slice of the lungs by the Ray,
{\em et al} \cite{Ray} method.} 
\label{Fig:1}
\end{figure}

\begin{figure}

\includegraphics[height=4.0cm,width=8.0cm,angle=0]{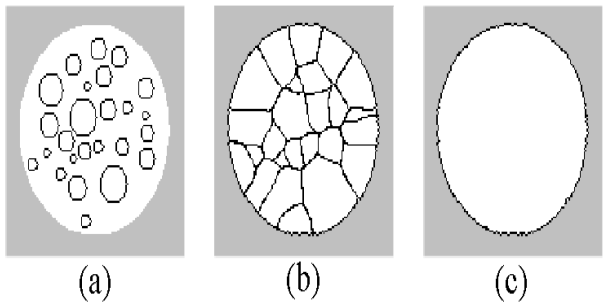}
\caption{Evolution of snakes (here circular, although one can use 
rectangular snakes as well) in a hypothetical circle image.}
\label{Fig:2}

\end{figure}

In order to map the merging snake scenario to statistical 
thermodynamical systems, we first provide the mathematical background 
for a parametric active contour or snake. A snake is a curve 
${\bf C}(s) = (p(s),q(s))$
defined by the parameter $s \in [0,1]$. The snake is evolved
in such a way that it minimizes the energy functional \cite{Xu}

\begin{equation}
E_s = \int_{0}^{1} (\frac{1}{2} \{\alpha {|{\bf C'}(s)|}^2 +
\beta {|{\bf C''}(s)|}^2 \} + E_{\rm ext} [{\bf C}(s)])\:\:ds
\end{equation}

\noindent
where the first two terms give the internal energy of the snake 
($\alpha,\:\:\beta \geq 0$) and $E_{\rm ext}$ represents the external 
energy added to the system. ${\bf C'}$ and ${\bf C''}$ are the first
and second derivatives of the snake with respect to $s$. 
An example of external force for the snake is the gradient force:
$E_{\rm{ext}} = -{|{\bf \nabla} I(x,y)|}^2 $, where $ I(x,y) $ is the
image pixel intensity. In general, 
in the presence of an external force ${\bf V}({\bf C}(s,t))$, the time 
dynamics of such a snake is given by \cite{Ray}

\begin{equation}
\frac{\partial {\bf C}(s,t)}{\partial t} = \alpha {\bf C''}(s,t) - 
\beta {\bf C''''}(s,t) + {\bf V}({\bf C}(s,t)). 
\end{equation}

\noindent
Note that the snake in eqn. (2) is now a function of time $t$ as 
well. The stationary solution of eqn. (2) corresponds to a snake that 
minimizes the energy functional. Ray {\em et al} proposed an external force
$ V(x,y) $ obtained by solving the following partial differential 
equation (PDE) applying Dirichlet boundary condition[1]:

\begin{equation}
g(|{\bf \nabla}f|) {\nabla}^2 {\bf V} - (1 - g(|{\bf \nabla}f|)) ({\bf V} - 
{\bf \nabla}f) = {\bf 0},
\end{equation} 

\noindent
where $g(\alpha) = \exp(-K \alpha)$ and $f(x,y) = -E_{\rm ext}(x,y)$, 
$K$ being a tunable
parameter controlling the smoothness of the external snake force field.
Here we study the resultant dynamics due to 
the evolution of a number of such snakes, defined by the above
system of forces. 

When multiple snakes are evolved inside the desired closed boundary, 
the growth algorithm confirms that they all finally 
merge into a single snake after a finite time. This {\em merging of
snakes} is basically a topological effect and naively the
phenomenology reminds one of "nucleation" as seen in classical first-order
thermodynamic systems \cite{nucleation}. In our case, we allow a finite
number of GGVF snakes, each with a finite starting radius, to evolve in a 2-D 
plane and then numerically evaluate a few measurables -- the 
{\em nucleation time (NT)}, the {\em bounded area (BA)} after nucleation has
occurred, the 
{\em critical radius (CR)} at the time of nucleation and also the
{\em nucleation rate (NR)}, all as functions of snake evolution time. The 
respective quantities are defined as follows -- 
{\em nucleation time} is the time required for all the snakes to merge
together as a single unit,
{\em bounded area} is the sum of the areas of the initial snakes {\em before}
complete merging and the area under the single snake {\em after} merging, 
{\em critical radius} is the radius of curvature of the merged
structure and
{\em nucleation rate} is the ratio of the number of snakes to the
bounding area, before the merging has actually taken 
place. One should note that by {\em complete merging} we refer to the critical
phase when all the initial snakes merge together for the first time.

We perform two numerical experiments with merging. In the first
experiment, we start with a circular binary image of
radius $R$ containing $N$ number of circular snakes, each of radius
$r<R$, randomly distributed inside. 
As described before, the initial snakes are driven by GGVF
forces and they maintain a non-overlapping dynamics. In another
numerical experiment, we vary the radii of the smaller circles and later also
vary their total number. Both numerics are repeated with
varying sizes of the initial domain $R$. The enclosed figures 3 and 4 
show the variation of the bounding area and the critical radius
against time respectively, in non-dimensionalized units, in loglog plots. 

\begin{figure}
\includegraphics[height=6.0cm,width=6.0cm,angle=0]{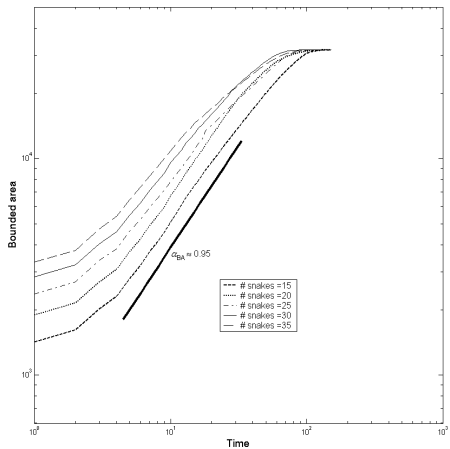}
\caption{Bounding area vs time in log-log scale.} 
\label{Fig:3}
\end{figure}

\begin{figure}
\includegraphics[height=6.0cm,width=6.0cm,angle=0]
{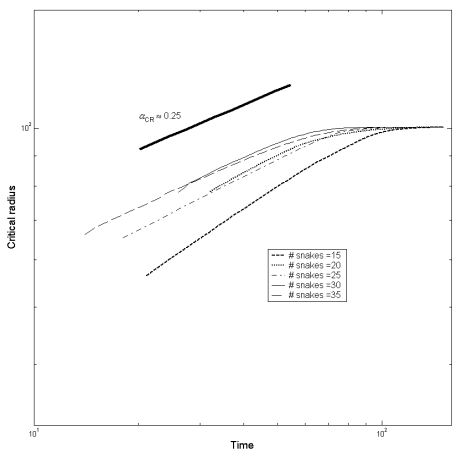}
\caption{Critical radius vs time (log-log), after complete merging of the 
snakes.}
\label{Fig:4}
\end{figure}

\noindent
We find that all the graphs show excellent scaling in a broad domain, with 
the nonlinear zones referring to the saturation limits in each case. In the
figures shown, the bounding image radius is 80 units while the 
corresponding starting snake radius is 5 units. We have simulated using 
different number of initial snakes, indicated in the legend of the figures. 
The plots show power law
variations of each of the measured quantities with time, {\em i. e.} $y
\sim x^{\alpha}$ and the two relevant exponents, that of {\em critical radius} 
and {\em bounding area}, have the values  $\alpha_{\rm BA}
\approx 0.9-1.0,\:\:\alpha_{\rm CR} = 0.25$. We have used multiple
combinations of the parameter values but the exponents remain {\em universal},
that is they invariably follow a power law statistics. 

To analyse the results further, we begin with the
hypothesis that the phenomenology of the whole process, that of various snakes
{\em merging} together under certain boundary conditions in the
presence of suitably defined external forces, is fundamentally similar
to that of a domain growth process as encountered in the realms of
classical nucleation. In what follows, we try to evaluate the time rate of
evolution of a system of snakes before and after all
the snakes have completely merged with each other. We 
start from a model of solutes trying to nucleate in a solution and then
compare the results with the merging mechanism of snakes we find in the
AC method.

Our starting description is that of the diffusive growth of two
dimensional spherical droplets (circles, in 2-D) in a
Lifshitz-Slyozov \cite{nucleation, LStheory, binderstauffer} type of
continuum theory. We start with the thermodynamic definition of the work 
$ \Delta W $ required to form a nucleus from an aggregate of solute particles 

\begin{equation}
\Delta W = \delta (E - T_0 S + P_0 V)
\end{equation}

\noindent
where $ \delta (E - T_0 S) $ is the increment in the free energy of the
system and $ \delta (P_0 V) $ is the associated external work done.
Optimizing this total work factor, one can show 
\cite{nucleation} that the critical radius of a nucleating system, this being 
the rate of merger of snakes in AC method, is given by

\begin{equation}
R_{\rm CR} = \frac{\beta s'}{\mu'(P) - \mu_0'(P)}
\end{equation}

\noindent
Here $\mu_0'$ and $s'$ are the chemical potential and molecular
surface area of the nucleus, $\beta$ is the surface tension and $\mu'$
is the chemical potential of the solute in the solution. As more and
more solutes start nucleating {\em i. e.} snakes merge, the solution
approaches the critical limit of supersaturation and in the steady
state, the {\em critical radius} grows as

\begin{equation}
\frac{dR(t)}{dt} = D (\frac{\partial c}{\partial r})|_{r=R}
\end{equation}

\noindent
where $c(R)$ represents the spherically symmetric concentration distribution 
of snakes around a nucleus of radius $a$, $D$ being the diffusion 
coefficient.  
Following the treatment of Lifshitz-Slyozov \cite{nucleation,LStheory}, one 
can show that if $R_{\rm CR}(0)$
is the critical radius at the beginning of the merging process, then for
a diffusive nature of relaxation, the radius follows a dynamics 

\begin{equation}
\frac{dR(t)}{dt} = \frac{{R_{\rm CR}^2(0)}}{R(t)} (\frac{1}{R_{\rm CR}(t)} - 
\frac{1}{R(t)})
\end{equation}

\noindent 
We now define the dimensionless quantities $x(t) = \frac{R_{\rm CR}(t)}
{R_{\rm CR}(0)}$, $u(t) = \frac{R(t)}{R_{\rm CR}(t)}$ and 
$\tau = 2 \log x(t)$. The last one increases
monotonically from 0 to $\infty$ as the time $t$ increases likewise. 
Combining (6) and (7) in 2D, the dynamics is given by  

\begin{equation} 
\frac{d u^2(\tau)}{d{\tau}} = \gamma (1 - \frac{1}{u}) - u^2
\end{equation}

\noindent
where $\gamma = \frac{dt}{R_{\rm CR}(0) dx} > 0$. A linear stability analysis 
of the above nonlinear equation gives a fixed point at $\gamma_0 = 
\frac{27}{4}$. We are interested in the dynamics around an
$\epsilon$-neighborhood of this point $\gamma_0$. If $\gamma(\tau)=
\frac{27}{4}[1 - {\epsilon}^2(\tau)]$ ($\epsilon \rightarrow 0$ as 
$\gamma \rightarrow \infty$), then near the critical point 
($ u_0 = {(\gamma/2)}^{1/3} = 3/2 $) the merging snakes follow a 
dynamics defined by   

\begin{equation}
\frac{du}{d{\tau}} = -2 {(u - \frac{3}{2})}^2 - \frac{3}{4} {\epsilon}^2
\end{equation}

\noindent
The time variation of the merging nuclei (snakes) is

\begin{equation}
x(t) = \frac{4}{27} \frac{\sqrt{t}}{R_{\rm CR}(0)}
\end{equation}

\noindent
Thus the bounding area ($\sim x^2(t)$) of the merging snakes
grow at the rate of $t$ which is roughly speaking our numerical estimate 
(0.9-1.0) also
(Fig. 3). However, the above analysis is only valid {\em before} complete
merging of the snakes has occurred. In the {\em merged} phase when 
the resultant  
asymmetrical structure continues growing finally to coalesce with the bounding
image, the system dynamics is modified. To analyse the situation,
we start from eqn.(7). The equation may alternatively be represented as

\begin{equation}    
\frac{dR(t)}{dt} = \frac{D \delta c(R(t))}{R(t)} \sim \frac{1}{T} \
\frac{1}{R^2(t)}.
\end{equation}

\noindent
$T$ in the above equation is the temperature of the solution in the
equivalent thermodynamic problem which in our case, is a measure of the 
average surface energy $E$, $ E \propto T $, of the system. 
Evidently, in 2D, $E$ goes as $2 \pi R \beta$. 
Plugging these values in the above
equation, we see that after complete merging of the snakes has taken place,
the effective radius of curvature of the resultant structure grows as 
$R_{\rm CR}(t) \sim t^{1/4}$ (Fig. 4). Once again our numerical result 
is in exact harmony with the theory.

Our above analysis, both numerical and analytical clearly suggests that below 
the apparently simplistic level of the GGVF application, the system dynamics 
has a more fundamental symmetry. This symmetry comes from  
the fact that the GGVF method lies in the same
universality class as that of a classical nucleation model. This first-order
critical response of the system is what 
provides the subtlety of the underlying physics. There is, however,
a notable shortcoming with the GGVF technique, that it does 
not allow us the liberty of starting with an initial condition at an 
arbitrary location. If the
snake starts at a position in which a major portion of the initial snake is
outside the desired boundary or {\em vice versa}, 
then the snake driven by GGVF will not converge to the
actual boundary. Although it can be shown by the Reed-Simon's
theorem \cite{reedsimon} that a convex set (a circle, say) growing within 
a larger convex set (a larger circle or rectangle, say) will always merge 
with the outer boundary under the action of isotropic driving forces, to 
the best of our knowledge, no such mathematical lemma exists for a
convex set growing in a concave set or vice versa. 

In summary, our achievement in the present paper
has been to analyze the physical foundation of the GGVF merging technique
and to provide answer to the rather puzzling question as to why it
works so accurately. In the process, we have shown that 
the answer lies in the general scaling behavior of the underlying 
nucleation dynamics, defined by proper scaling laws. This clearly 
indicates that an active contour system is in the same universality class as the
nucleation model we considered. Our results, in unison with probable
biomedical applications, are expected to inspire further 
studies in the understanding of lung-based diseases.


\begin{thebibliography}{99}



\bibitem{Ray} N. Ray, S. T. Acton, T. Altes, E. E. de Lange and 
J. R. Brookeman, IEEE Trans. Med. Imaging {\bf 22}, 189 (2003). 

\bibitem{Xu} C. Xu and J. L. Prince, Signal Processing {\bf 71}, 
131 (1998).

\bibitem{Kass} M. Kass, A. Witkin and D. Terzopolous, Int. J. Comput. Vis.
{\bf 1}, 321 (1987).



\bibitem{Canny} J. F. Canny, IEEE Trans. Pattern Analysis and Machine 
Intelligence, 679 (1986).

\bibitem{McInerney} T. McInerney and D. Terzopolous, 
Med. Image Anal. {\bf 4}, 73 (2000).




\bibitem{nucleation} J. D. Gunton, M. San Miguel and P. S. Sahni in
{\em Phase transitions and critical phenomena}, edtd. by C. Domb and
J. L. Lebowitz, vol. 8, Academic Press (1983); E. M. Lifshitz and
L. P. Pitaevskii in {\em Physical Kinetics}, Landau and Lifshitz
Course of Theoretical Physics, vol. 10, Butterworth-Heinemann, 1981.

\bibitem{LStheory} I. M. Lifshitz and V. V. Sloyozov, J. Phys. Chem
  Solids {\bf 19}, 35 (1961); C. Wagner, Z. Electrochem. {\bf 65}, 581
(1961).

\bibitem{binderstauffer} K. Binder and D. Stauffer, Adv. Phys. 
{\bf 25}, 343 (1976).

\bibitem{Langer69} J. S. Langer, Ann. Phys. {\bf 41}, 108 (1967);
Ann. Phys. {\bf 54}, 258 (1969).

\bibitem{Langer73} J. S. Langer and L. A. Turski, Phys. Rev. A {\bf 8}, 
3230 (1973); {\em ibid}, Phys. Rev. A {\bf 22}, 2189 (1980).

\bibitem{reedsimon} M. Reed, B. Simon and S. M. Reed in the "Methods of
Modern Mathematical Physics: Fourier Analysis, Self-adjointness", Academic
Press, 1997.

\end{thebibliography}
\end{document}